\documentclass[10pt,aps,prb,twocolumn,showpacs,amssymb]{revtex4-1}
\usepackage{amsmath,graphics,epsfig,mathrsfs}
\usepackage{bm}
\usepackage{epsfig}
\begin{document}

\title{Spin Transfer Torque in Fully Insulating Magnetic Tunnel Junctions}

\author{Y. Yuan and A. Manchon}
\affiliation{King Abdullah University of Science and Technology (KAUST),
Physical Science and Engineering Division, Thuwal 23955-6900, Saudi Arabia}
\begin{abstract}
Voltage-driven spin transfer torque in a magnetic tunnel junction comprising magnetic insulating electrodes is studied theoretically. In contrast with the conventional magnetic tunnel junctions comprising transition metal ferromagnets, the spin transfer torque presents unconventional bias dependencies, related to the presence of spin-dependent Fowler-Nordheim tunneling processes. In particular, we find that (i) the out-of-plane torque generally dominates the in-plane torque, (ii) out-of-plane torque and in-plane torque are symmetric and asymmetric at low bias voltage, respectively, and (iii) both of torques show a dramatic enhancement at large bias voltage. Materials consideration are discussed and we show that due to the low damping parameter expected in magnetic insulators a spin transfer torque can be experimentally observed in such systems.
\end{abstract}
\date{\today}
\maketitle

%\clearpage

\section{Introduction}

Ferromagnetic and ferrimagnetic insulators have gained increasing interest in the recent years with the observation of thermally-induced pure spin currents attributed to the so-called spin Seebeck effect \cite{seebeck}. These magnetic materials possess a very low magnetic damping parameter ($\alpha\approx$6.7$10^{-5}$ for Y$_3$Fe$_5$O$_{12}$) resulting from the absence of conduction electrons and constitute ideal tools to investigate spin waves propagation and non-linear properties over long distances \cite{hillebrand}. Indeed, the electrical generation of spin-waves in Y$_3$Fe$_5$O$_12$\cite{Kajiwara,wang} (YIG) through spin transfer torque \cite{Slon1996} as well as the observation of spin pumping\cite{tserko} from YIG into Au and Pt layers \cite{heinrich,sandweg} have proven that these complex materials may be relevant for spin transfer torque and spin pumping studies. Whereas Slonczewski proposed that magnons thermally excited in a magnetic insulator can exert a torque on an adjacent free layer\cite{slonc2010}, Xiao et al. \cite{xiao} were able to interpret spin Seebeck experiments \cite{seebeck} in terms of spin pumping induced by thermal magnons.\par

Besides their low magnetic damping constant, magnetic insulators are also expected to make a difference in the race for the largest tunneling magnetoresistance (TMR) effect in magnetic tunnel junctions (MTJs) composed of two ferromagnets separated by an insulator \cite{moodera95}. While pioneering studies used amorphous Al$_2$O$_3$ (alumina) barriers and transition metal electrodes \cite{moodera95}, an outstanding breakthrough was achieved ten years ago with the discovery of electron wave filtering in Fe(100)/MgO crystalline systems \cite{mgoth,yuasa} where TMR up to 604\% for a regular MTJ \cite{ikeda:082508} and 1056\% for a double MTJ \cite{jiangmgo} were obtained. In the case of magnetic insulators, since the barrier height is spin-dependent, the current polarization increases with the barrier thickness\cite{spinfilter} $P\approx\tanh qd\frac{\Delta}{U}$. Here $d$, $q$, $U$ and $\Delta$ are the barrier thickness, evanescent wave vector, barrier height and spin splitting of the magnetic insulator. Therefore, magnetic insulators have also been used as spin filters in quasi-spin valves structures. In these structures, the polarizer consists of ferro- or ferrimagnetic tunnel barriers such as EuS \cite{moodera88,nagahama,leclair}, NiFe$_2$O$_4$ \cite{NiFe2O4}, CoFe$_2$O$_4$ \cite{CoFe2O4}, BiMnO$_3$ \cite{BiMnO3} and EuO \cite{SantosEuO}, the spacer is either tunneling (AlOx) or metallic and the analyzer is either a half-metal (LSMO), a transition metal ferromagnet or even another magnetic insulator \cite{Miao}. The magnetic insulators directly filter the tunneling electron spin and provide large TMR ratio at low temperature, up to 60\% at 1K in the case of fully insulating MTJs \cite{Miao}.

The mechanisms of spin-dependent tunneling through such structures have been investigated experimentally and theoretically unraveling complex bias dependence of the spin transport such as a spin-selective Fowler-Nordheim\cite{Fowler} (FN) tunneling processes \cite{nagahama,Miao}: At large enough bias, the majority spin sees a reduced barrier thickness (FN tunneling) while the minority spin are still in the direct tunneling regime. The onset of the spin-selective FN process results in an effective enhancement of the TMR \cite{Miao}, in sharp contrast with the monotonous decrease observed in regular MTJs \cite{reviewtsymbal}.\par

In the context of the recent observation of spin transfer torque in YIG \cite{Kajiwara,wang}, a few theoretical studies have addressed the nature of the spin torque exerted on magnetic insulators. Inoue \cite{inoue} derived the analytical expression of the electrically driven spin torque between a ferromagnetic metal and a magnetic insulator and Jia et al. \cite{jia} calculated the mixing conductance at the interface between YIG and Ag using Density Function Theory. These two theoretical studies intend to explain the observed magnitude of the spin torque induced by spin Hall effect observed in Refs. \onlinecite{Kajiwara,wang}. They show that although spin torque is small in such structures it can still be detected through spin wave measurements. In the present study, we focus our attention on the fully insulating MTJ proposed by Miao et al. \cite{Miao}. In particular, whereas Refs. \onlinecite{Kajiwara,wang} measured the spin torque in the linear transport regime (low bias), such fully insulating MTJ display strong non-linearities at large biases, illustrated by the dramatic enhancement of the TMR. This enhancement indicates a regime where the spin polarization is maximum and therefore one can reasonably expect a large spin transfer torque efficiency. We show that the spin torque is dominated by the out-of-plane torque and can be experimentally exploited in current-driven magnetization dynamics studies. In Section \ref{sec:II}, we discuss the nature of the spin torque in magnetic insulators using a simple bilayer model. The theoretical model for the fully insulating MTJ is exposed in Section \ref{sec:III} and numerical results are reported on Section \ref{sec:IV}. Materials consideration and experimental relevance are discussed in Section \ref{sec:V}. The conclusion is given in Section \ref{sec:VI}.\par

\section{Bilayer Model\label{sec:II}}

Before entering in the details of the device physics, let us turn our attention to the spin transport taking place at the interface between two ferromagnets. This model follows the spirit of Stiles and Zangwill \cite{stiles2002} seminal work on spin torque in metallic spin-valves. Our system consists of two adjacent semi-infinite (either metallic or insulating) ferromagnets whose magnetizations are misaligned by an angle $\theta$. Considering a majority spin-polarized electron ($\uparrow$) flowing from left to right, its free-electron wave function reads
\begin{eqnarray}
&&\psi(z<0)=(e^{ik_L^\uparrow z}+R_\uparrow e^{-ik_L^\uparrow z})|\Uparrow\rangle+R_\downarrow e^{-ik_L^\downarrow z}|\Downarrow\rangle\\
&&\psi(z>0)=T_\uparrow e^{ik_R^\uparrow z}|\uparrow\rangle+T_\downarrow e^{ik_R^\downarrow z}|\downarrow\rangle\label{eq:ins}
\end{eqnarray}
where $|\Uparrow\rangle=\cos\frac{\theta}{2}|\uparrow\rangle+\sin\frac{\theta}{2}|\downarrow\rangle$ and $|\Downarrow\rangle=\sin\frac{\theta}{2}|\uparrow\rangle-\cos\frac{\theta}{2}|\downarrow\rangle$. $R_{\uparrow,\downarrow}$ ($T_{\uparrow,\downarrow}$) is the spin-dependent reflection (transmission) coefficient and $k_i^{\uparrow,\downarrow}$ are the spin-dependent wavevectors in the $i$th layer. Note that these wavefunctions are expressed in the quantization axis of the right layer. The solution of the system is obtained by a standard wave matching procedure throughout the junction. The applied bias voltage $V$ drives a (conserved) charge current $J_e$ and (spatially dependent) spin current ${\cal J}_{s}$ through the junction. The charge and spin current densities carried by one electron described by the wave function $\psi$ are given by
\begin{eqnarray}
&&J_{e}=\frac{e\hbar}{m}\Im(\psi^{\ast}\nabla\psi)\\
&&{\cal J}=\frac{\hbar}{m}\Im(\psi^{\ast}{\hat{\bm\sigma}}\otimes\nabla\psi)
\end{eqnarray}
where ${\hat{\bm\sigma}}=({\hat\sigma}_{x},{\hat\sigma}_{y},{\hat\sigma}_{z})$ is the vector of Pauli spin matrices. In the absence of spin-flip scattering, since the total angular momentum is conserved, the spin current lost is transferred to the local magnetic moment: the spin torque is directly related to the transverse spin current components ${\cal J}_{||}+i{\cal J}_{\bot}$ giving rise to the in-plane and perpendicular torques, respectively. In the case of an interface between two metallic ferromagnets, one obtains the transverse spin current impinging on the right layer
\begin{eqnarray}
&&{\cal J}_{||}^{z>0}=\frac{4}{\hbar}\frac{k_L^2(k_R^\uparrow+k_{R}^{\downarrow})}{(k_L+k_{R}^{\uparrow})^2}\Re[e^{i(k_{R}^{\uparrow}-k_{R}^{\downarrow})z}]\sin\theta\\
&&{\cal J}_{\bot}^{z>0}=-\frac{4}{\hbar}\frac{k_L^2(k_R^\uparrow+k_{R}^{\downarrow})}{(k_L+k_{R}^{\uparrow})^2}\Im[e^{i(k_{R}^{\uparrow}-k_{R}^{\downarrow})z}]\sin\theta
\end{eqnarray}
To obtain the actual spin torque exerted on the right layer, the above expressions have to be averaged over the Fermi surface. In the case of metallic interfaces, it turns out that the perpendicular torque vanishes (a detailed discussion can be found in Ref. \onlinecite{stiles2002}). At this stage, we point out that, due to the coherent interference of precessing spins, the torque extends in the right ferromagnet on the distance $\lambda_\varphi\approx 2\pi/(k_R^\uparrow-k_R^\downarrow)$ \cite{stiles2002}. When the right layer becomes half-metallic, minority electrons can not propagate and an exponentially damped oscillating torque is found \cite{manchon2008}.\par
 
The situation is quite different when considering the interface between a metallic ferromagnet and a insulating ferromagnet. In this case the wavevectors in Eq. \ref{eq:ins} should be replaced by $k_{R}^{\uparrow,\downarrow}\rightarrow iq_{R}^{\uparrow,\downarrow}$. The spin current reads
\begin{eqnarray}
&&{\cal J}_{||}^{z>0}=\frac{4}{\hbar}\frac{k_L^3(q_R^\uparrow+q_R^\downarrow)^2}{(k_L^2+q_R^{\uparrow2})(k_L^2+q_R^{\downarrow2})}e^{-(q_R^{\uparrow}+q_R^{\downarrow})z}\sin\theta\\
&&{\cal J}_{\bot}^{z>0}=-\frac{4}{\hbar}\frac{k_L^2(k_L^2+q_R^\uparrow q_R^\downarrow)(q_R^\uparrow-q_R^\downarrow)}{(k_L^2+q_R^{\uparrow2})(k_L^2+q_R^{\downarrow2})}e^{-(q_R^{\uparrow}+q_R^{\downarrow})z}\sin\theta
\end{eqnarray}
Since there are only evanescent states in the barrier, the torque exponentially decreases away from the interface. Interestingly, the perpendicular torque is now proportional to $q_R^\uparrow-q_R^\downarrow$ which indicates that it should be smaller than the in-plane torque. Finally, in the case of two insulators, we obtain
\begin{eqnarray}
&&{\cal J}_{||}^{z>0}=0\\
&&{\cal J}_{\bot}^{z>0}=-\frac{4}{\hbar}\frac{q_L^2(q_R^\uparrow-q_R^\downarrow)}{(q_L+q_R^{\downarrow})^2}e^{-(q_R^{\uparrow}+q_R^{\downarrow})z}\sin\theta
\end{eqnarray}
Interestingly, there is simply no in-plane torque in this system and only a field-like perpendicular torque survives. This can be understood by noticing that in the absence of propagating states, no spin precession occurs in the system. Therefore, the right layer feels an effective evanescent spin density aligned along ${\bf z}$, which acts as an effective field. The more realistic case of a fully insulating MTJ, comprising two magnetic insulators separated by a normal insulator, is more complex. However, the present simple interfacial models indicate that a dominant field-like torque can be reasonably expected in such systems.

\section{Full Junction Model\label{sec:III}}

\begin{figure}
\centering
\includegraphics[scale=0.8]{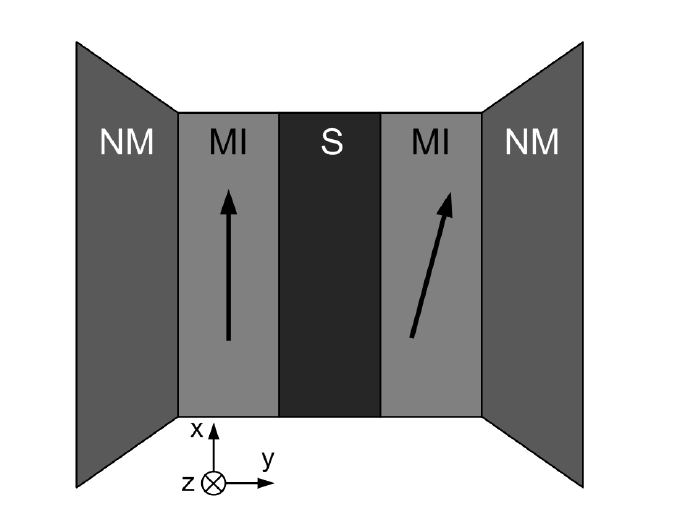}
\caption{\label{fig:fig1}Schematics of the fully insulating MTJ. NM, MI, and S represent the non-magnetic electrodes,
the magnetic insulators, and the insulating spacer, respectively.}
\end{figure}

We consider a fully insulating magnetic tunnel junction as sketched in Fig. \ref{fig:fig1} in which two normal metals (NM) work as electrodes and two magnetic insulating layers (MI) are separated by one non-magnetic insulating spacer (S). The magnetization of the left magnetic insulator ${\bf M}_L$ is aligned along ${\bf z}$ and the one of the right magnetic insulator ${\bf M}_R$ lies in the x-z plane and is rotated by an angle $\theta$ with respect to ${\bf M}_L$.

\begin{flushleft}
\begin{table}
\begin{tabular}{c|c|c|c|c|c|c|}
 &EuS&EuO&BiMnO$_3$&NiFe$_2$O$_4$&CoFe$_2$O$_4$&Y$_3$Fe$_5$O$_{12}$\\
  & Ferro & Ferro & Ferro & Ferri & Ferri & Ferri \\ \hline\hline
$T_c$ (K) & 16.6 & 69.3 & 105 & 850 & 796 & 550\\ \hline
  U (eV) & 1.65 & 1.12 & 1.65 & 0.98-1.2 & 0.57-0.8 & 1.4 \\ \hline
  2$\Delta$ (eV) & 0.54 & 0.36 & 2.3 & 1.21 & 1.28 & 0.33\\ \hline
    Reference & \onlinecite{moodera88} & \onlinecite{SantosEuO} &\onlinecite{Shishidou} & \onlinecite{Szotek} & \onlinecite{Szotek} &\onlinecite{jia}\\ \hline
\end{tabular}
\caption{\label{tab1} Curie temperature ($T_c$), energy gap (U) and exchange splitting (2$\Delta$) of several ferro- and ferrimagnetic insulators.}
\end{table}
\end{flushleft}

Magnetic insulators are present under various crystal structures such as rocksalt (EuS, EuO), perovskite (BiMnO$_3$), inverse spinel (NiFe$_2$O$_4$,CoFe$_2$O$_4$) or garnet (Y$_3$Fe$_5$O$_12$) and their Curie temperature varies over a wide range (16.6-850K). They display a gap of the order of $U\approx$1$\pm$0.5 eV and their conduction band shows spin splitting of the order of $\Delta\approx0.3-2$eV (see Table \ref{tab1}). In the present model, the band structure of these materials is represented by a simple free electron model. Although important features such as interfacial resonant states and electron wave filtering might have some impact on the tunneling spin transport (as in MgO/Fe(100) systems \cite{mgoth}), we disregard these effects at this stage and retain only a minimal set of parameters. The barriers height, thickness and splitting of the insulators are denoted U$_{L,S,R}$, $d_{L,S,R}$, and $\Delta_{L,R}$ respectively. Note that the spacer is a non-magnetic insulator ($\Delta_S=0$). Under the free electron approximation, the Hamiltonian of the system reads
\begin{equation}\label{eq:ham}
H=-\frac{\hbar^{2}}{2m^{\ast}}\frac{\partial^{2}}{\partial x ^{2}}+U_{L,R}+\Delta_{L,R}{\hat{\bm \sigma}}\cdot{\bf M}_{L,R}+\frac{eV}{2}+E_F-\frac{x}{d}eV
\end{equation}
where $d$ is the total thickness of the composite barrier, $m^{\ast}$ is effective mass, and $E_F$ is Fermi energy in the metallic electrode. The solution of Eq. (\ref{eq:ham}) is on the form
\begin{equation}\label{eq:hai}
|\psi\rangle=\sum_{\sigma}[aA_i(Z^\sigma(x))+bB_i(Z^\sigma(x))]|\sigma\rangle
\end{equation}
where Ai and Bi are the homogeneous Airy functions, and the argument $Z^\sigma(x)$ reads
%\begin{widetext}
\begin{eqnarray}
Z^{\sigma}(x)&=&\xi\left(\frac{2m^{\ast}}{\hbar^{2}}\left(U-\sigma \Delta+\frac{eV}{2}-\frac{eVx}{d}-E\right)+\kappa^{2}\right),\nonumber\\
\xi&=&\left(\frac{d\hbar^{2}}{m^{\ast}\mid eV\mid}\right)^{2/3}
\end{eqnarray}
%\end{widetext}
where $E,\kappa$ and $\sigma=\pm$ are the energy, the in-plane wave vector and spin projection of the incident electron. Previous methods such as Refs. \onlinecite{nagahama,Miao} have used Wentzel–Kramers–Brillouin (WKB) approximation instead of the exact Airy functions to solve the eigenstates of Eq. (\ref{eq:ham}). The WKB approximation is justified in the case of thick barriers only and does not capture the details of the Fowler-Nordheim resonances when the applied bias voltage exceeds the barrier height. On the other hand, whereas Airy functions are the exact solutions of the free electron Hamiltonian, they are numerically stable for only thin enough barriers. Thus, the gain in accuracy in describing the physical tunneling process at large biases is balanced by the constraint on the barrier thickness and height. Therefore, in the present study we choose to use the exact solutions provided by Airy functions but to apply this method to thin barriers only. Although the thickness of the barriers studied are smaller than the one usually attainable in the experiments, we believe that the physics addressed in this article is better described by the present method and will not significantly change for thicker barriers.\par

The spin-polarized current flowing through the structure exerts a local spin transfer torque (STT) \cite{Slon1996,Ralph} on the magnetic layers. The torque estimated in the left MI layer per unit area is equal to
\begin{eqnarray}
T&\equiv& -\int \nabla \cdot{\cal J}_{s}= {\cal J}_{s}(d_L)-{\cal J}_{s}(0)\\
&=&T_{||}{\bm M}_L\times({\bm M}_L\times{\bm M}_R)+T_{\bot}{\bm M}_L\times{\bm M}_R
\end{eqnarray}
where $T_{||}$ is the in-plane torque and $T_{\bot}$ is the out-of-plane torque.

%%%%%%%%%%%%%%%%%%%%%%%%%%%%%%%%%%%%%%%%%%%%%%%%%%%%%%%%%%%%%%%%%%%%%%%%%%%%%%%%%%%%%%%%%%%%%%%%%%
%%%%%%%%%%%%%%%%%%%%%%%%%%%%%%%%%%%%%%%%%%%%%%%%%%%%%%%%%%%%%%%%%%%%%%%%%%%%%%%%%%%%%%%%%%%%%%%%%%

\section{Results\label{sec:IV}}

Unless stated otherwise, the parameters for the system are: $U_{L,R}$= 0.8 eV, $\Delta_{L,R}$= 0.1 eV and $U_S$= 1 eV. In our definition, at positive bias, electrons travel from the left electrode to the right electrode.
\subsection{Tunneling Magnetoresistance }

\begin{figure}[hbtp]
\includegraphics[scale=1.2]{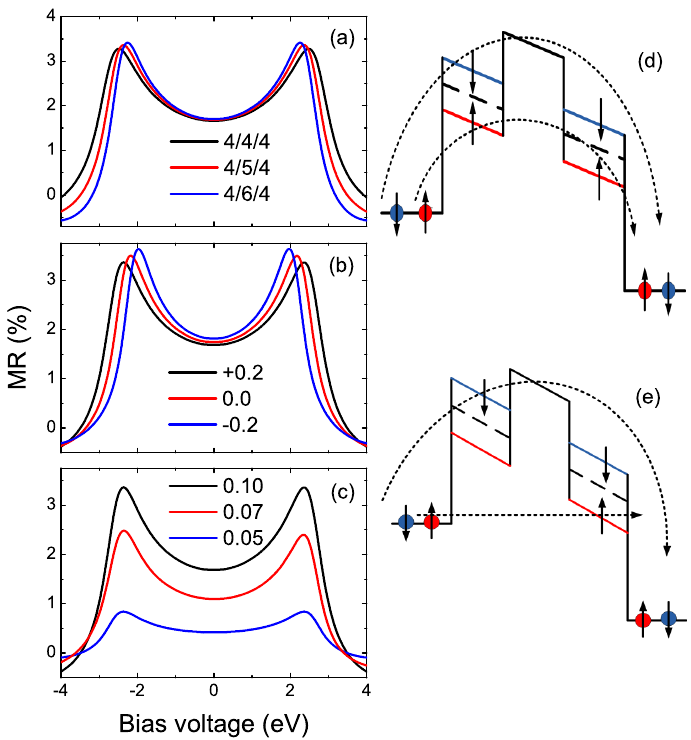}
\caption{Left panel: Bias dependence of the TMR for (a) different spacer thicknesses (b) different spacer barrier height and (c) different exchange splitting. Right panel: Energy profile of the fully insulating magnetic tunnel junction: (d) A direct tunneling occurs when the applied voltage is less than the lowest spin-up barrier height; (e) The Fowler-Nordheim tunneling is observed when the bias voltage exceeds the lowest spin-up barrier height \cite{Fowler}.}
\end{figure}

The bias dependence of the TMR, defined as as TMR=$(J_P-J_{AP})/J_{AP}$ (where P and AP refer to the parallel and antiparallel states, respectively), is represented in Fig.2 (a)-(c). As observed by Miao et al. \cite{Miao}, the TMR displays peaks at both positive and negative bias polarities. The feature has been identified by Miao \textit{et al.} \cite{Miao} as a direct consequence of so-called Fowler-Nordheim resonant tunneling \cite{Fowler}. When the applied bias voltage exceeds the barrier height for the majority spins, the effective thickness of the potential barrier seen by these electrons is reduced whereas it is unchanged for the minority spins. Therefore, the current for majority spins dramatically increases resulting in a significant enhancement of the TMR [see Fig.2(d,e)]. Fig.2(c) illustrates this effect for different exchange splitting $\Delta$. As expected, the effect is stronger for larger exchange splitting, due to a stronger spin filtering. Similarly, increasing the spacer thickness [Fig.2(a)] or decreasing the spacer barrier height [Fig.2(b)] enhances the TMR peak which occurs at smaller bias.\par

Interestingly, at larger bias a reversal of the TMR is observed as shown in Fig.6(a). When further increasing the bias voltage, the TMR ratio exhibits a clear oscillatory behavior and the period of the oscillations increase with the bias voltage. This oscillation has also been obtained by Montaigne et al. \cite{montaigne} is a regular MTJ when using an exact Airy description of the tunneling transport. This is directly related to the resonances associated with the FN process. Note that these oscillations are not captured by WKB approximation \cite{Miao,miaojap}.

%%%%%%%%%%%%%%%%%%%%%%%%%%%%%%%%
\subsection{Spin Transfer Torque}
To investigate the nature of the spin transfer torque, we set the magnetizations in the perpendicular configuration, $(\theta=\pi/2)$. The bias dependence at low voltage for both components of the spin torque is presented in Fig. 3. The form of the spin torque is\\
\begin{eqnarray}\label{Tip}
&&T_{\parallel}=a_{1}V+a_{2}V^{2}+a_{3}V^{3}\\
&&T_{\perp}=b_0+b_{1}V+b_{2}V^{2}\label{Top}
\end{eqnarray}
The in-plane torque is essentially antisymmetric in bias voltage (note a small asymmetry at large bias) whereas the out-of-plane torque is essentially symmetric presenting a quadratic bias dependence (with, again, a small deviation at large bias). This is in sharp contrast with the bias dependence of the torque obtained in regular symmetric magnetic tunnel junctions \cite{Theodonis,manchon2008,Heiliger}. For comparison, Theodonis \textit{et al.} \cite{Theodonis} found a dependence on the form
\begin{eqnarray}
&&T_{\parallel}=a_{1}V+a_{2}V^{2}\\
&&T_{\perp}=b_0+b_{2}V^{2}
\end{eqnarray}
Furthermore, whereas the out-of-plane torque is found to be smaller than the in-plane torque in conventional tunnel junctions \cite{Theodonis}, we find that it dominates the spin torque in fully insulating junctions. The influence of the barrier thickness and height are reported in Fig.3 (a,c) and (b,d), respectively. The magnitude of both in-plane and out-plane torques decrease when increasing the spacer thickness or height, as expected. The torque magnitude as a function of the exchange are plotted in Fig.3 (e,f).

\begin{figure}[hbtp]
\includegraphics[scale=1.2]{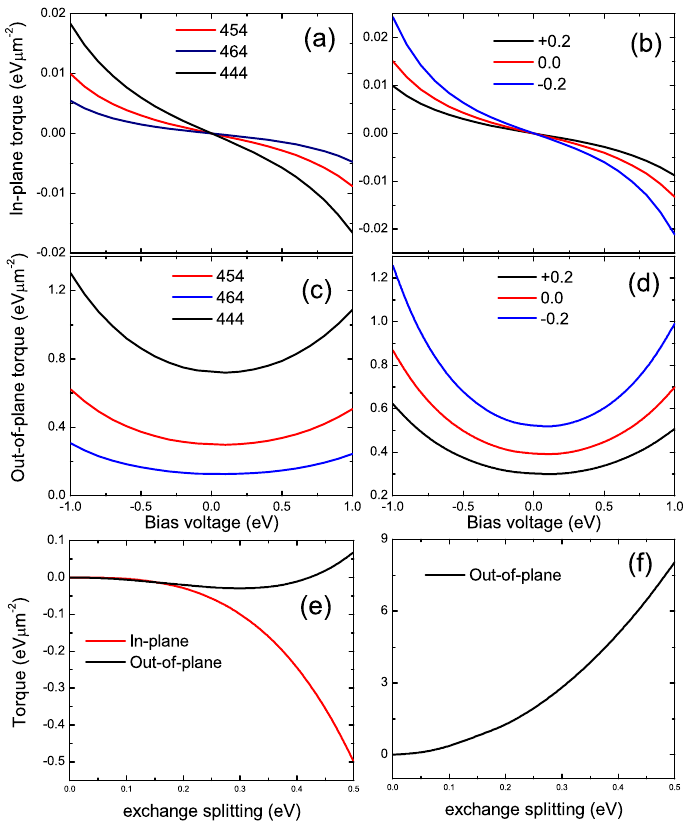}
\caption{Low bias dependence of (a,d) $T_{||}$ and (b,d) $T_\bot$ components of the torque exerted on left MI layer in a symmetric junction for $\theta=\pi/2$ together with the corresponding effects of different spacer thicknesses (a,c), different spacer barrier heights (b,d) and different exchange splitting (e,f).}
\end{figure}

At large biases, the spin torque shows an dramatic increase by 2 to 3 orders of magnitude, as displayed in Fig. 4. Most importantly, whereas the out-of-plane torque is smaller than the in-plane torque at small biases (see Fig. 3), it largely exceeds the in-plane torque at large biases. The contribution of majority and minority electrons originating from left and right reservoirs to the spin torque has been calculated in Fig.5. Whereas the torque originated from majority and minority spins compete with each other to produce the effective in-plane torque, they add up to produce the out-of-plane torque. In particular, note that the resulting spin torque is more efficient at negative bias, which corresponds to a direct torque, the electrons flowing from the right electrode to the left one.

\begin{figure}[hbtp]
\includegraphics[scale=1.2]{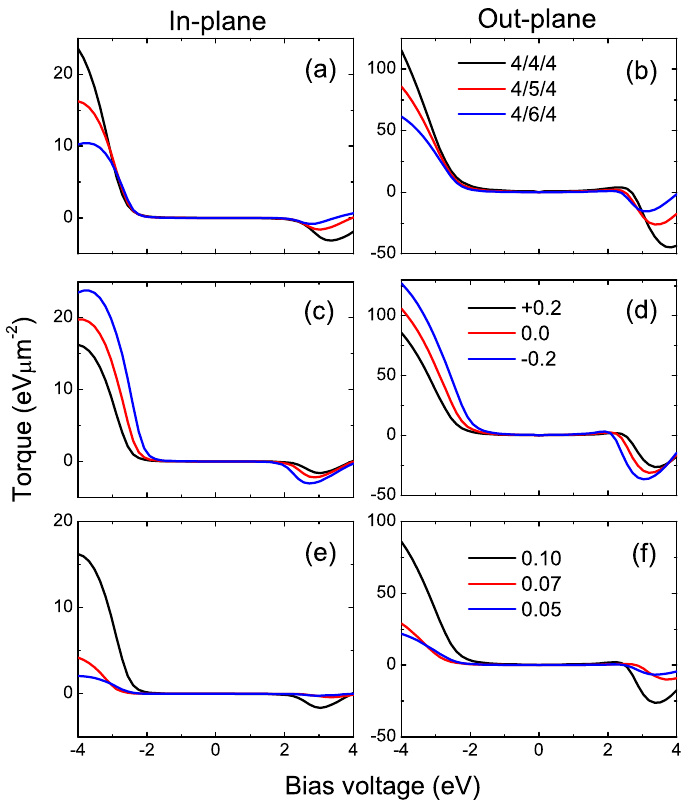}
\caption{High bias dependence of (a,b,c) $T_{||}$ and (d,e,f) $T_\bot$, together with the corresponding effects of (a,d) different spacer thicknesses (b,e) different spacer barrier heights (c,f) and different exchange splitting.}
\end{figure}

Mirroring the TMR behavior, the spin transfer torque at very large bias displays an oscillating bias dependence as shown in Fig.6(b,c), somewhat similar to the one observed by Tang \textit{et al.}\cite{Tang} in regular MTJs. This oscillation is attributed to the resonant states appearing when the barrier height goes below the Fermi energy of the incoming electrons, as mentioned above.

\begin{figure}[hbtp]
\includegraphics[scale=0.8]{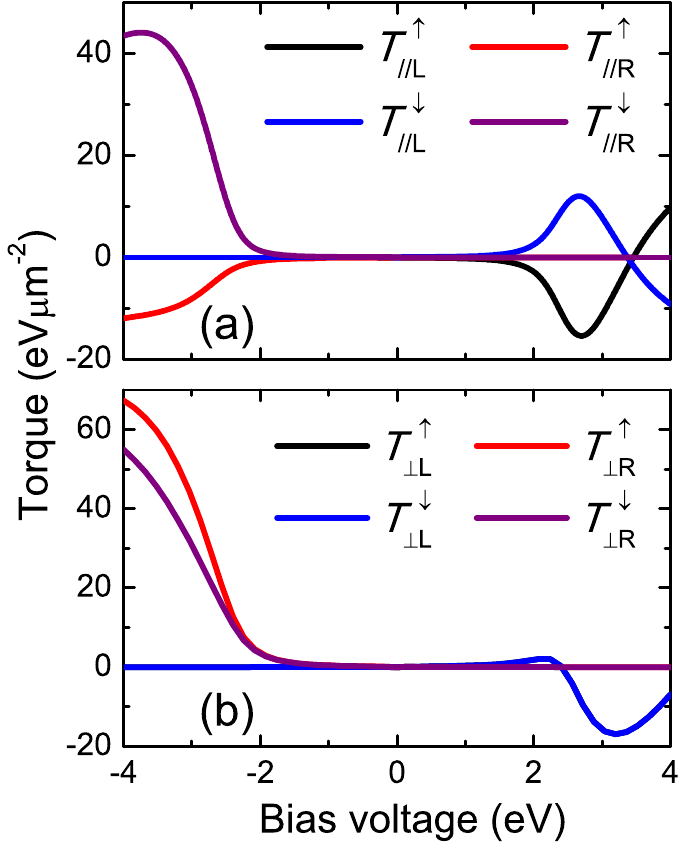}
\caption{Spin-dependent and reservoir-dependent contributions to the in-plane (a) and out-of-plane torques (b), exerted on the left MI layer.}
\end{figure}

\begin{figure}[hbtp]
\includegraphics[scale=1]{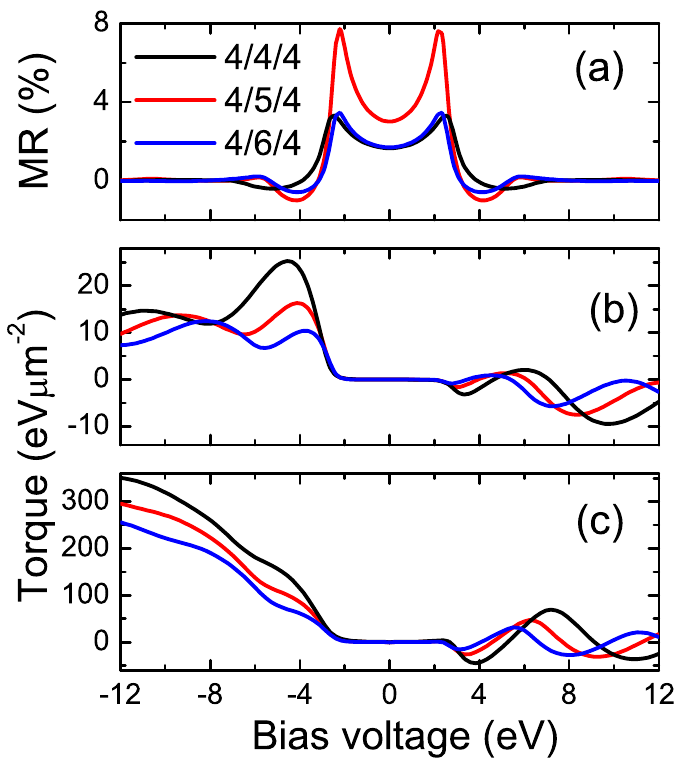}
\caption{Bias dependence of the TMR (a), in-plane (b) and out-of-plane torques (a) at very large biases for different spacer thicknesses.}
\end{figure}
%%%%%%%%%%%%%%%%%%%%%%%%%%%%%%%%%%%%%%%%%%%%%%%%%%

\section{Discussion\label{sec:V}}

The magnitude of the spin transfer torque calculated using the present model is  $\approx$0.5 $eV.\mu m^{-2} $ at 1V, which is about two orders of magnitude smaller than what has been calculated in regular MTJs using the same free electron approach (30-100 $ eV \mu m^{-2} $ at 1V in Ref. \onlinecite{wil}). However, in the present discussion we show that in principle, this difference does not prevent from observing a measurable effect of the current on the magnetization dynamics.\par

First, we have deliberately chosen conservative materials parameters, such as a small exchange energy $\Delta=0.1$eV. As shown in Fig. 3(e), the torque is more-than-linear as a function of the exchange and therefore, in realistic materials a larger spin torque is expected (see Table I). From the materials viewpoint, it is necessary to search for compounds with low barrier height, strong spin splitting and large Curie temperature (to favor thermal activation of the magnetization dynamics). Following Table \ref{tab1}, CoFe$_2$O$_4$ might be an interesting candidate. However, a hurdle remains the thin layer constraint. Miao et al.\cite{Miao} were able to grow 0.6 nm thick of EuS layer by thermal evaporation, which renders spin transfer studies reachable. Furthermore, the AlOx spacer could be replaced by a metallic spacer. In such a structure, resonant states would develop in the metallic spacer that could enhance the torque efficiency.\par

The fully insulating MTJ structure studied in this work is relevant for two types of experimental configuration. The first one is the usual bias voltage-driven magnetization switching such as the one currently observed in regular MTJs \cite{Huai}. In this configuration, when the magnetization of the layers lies in the plane, the intrinsic critical switching voltage (at zero temperature) is defined $V_c=\pm\frac{\alpha d}{a_1}(H_K+H_d)$ \cite{Ikeda}, where $d$ is the layer thickness, $H_K$ and $H_d$ are the anisotropy and demagnetizing fields and $a_1$ is the spin torque efficiency, defined in Eq. (\ref{Tip}). Another configuration is the bias voltage-driven domain wall motion, as proposed in Ref. \onlinecite{Chanthbouala}. In this case, the spin transfer torque is used to move a domain wall contained in the free layer. Whereas the domain wall depinning is governed by the in-plane torque, the domain wall steady state velocity is governed by the out-of-plane torque $v\approx-\frac{T_\bot}{\alpha}\Delta$. The interesting feature is that both quantities, magnetization switching threshold and domain wall velocity, are governed by the {\em ratio} between the spin torque and the damping constant $\alpha$. As mentioned in the introduction, since electron-magnon interactions are quenched in magnetic insulators, the damping is usually about much smaller in magnetic insulators than in transition metals ($\alpha\approx 6.7\times 10^{-5}$ for YIG compared to 0.01 for CoFe). Consequently, one can expect that the efficiency of the spin torque needs not to be as large as in MgO-based MTJs to observe the current-driven magnetization dynamics. Following our crude estimations, a bias voltage of 1 V should be sufficient to observe the magnetization dynamics in the case of a damping of $\alpha\approx10^{-4}$.

It is however important to notice that the extremely small magnetic damping of YIG is observed only for thick layers ($>$60 nm). Decreasing the layer thickness results in an increase of interfacial spin pumping effect \cite{tserko}  (and therefore an increase of the effective damping), as well as a modification of the band structure of the magnetic insulator \cite{notling} (associated with a decrease of the Curie temperature). Finally, YIG is known for its vanishing damping and stronger ferromagnets such as EuO or CoFe$_2$O$_4$ have generally larger damping factors. This may result in weaker current-driven magnetization dynamics which can still be detected using methods such as spin diode and spin-FMR measurements \cite{Sankey}.\par

\section{Conclusion\label{sec:VI}}
The current-driven spin torque in fully insulating tunnel junctions composed of ferromagnetic insulating electrodes has been studied theoretically using a free electron description. We find that: (i) the out-of-plane torque and in-plane torque are symmetric and antisymmetric at low bias voltage, respectively (ii) both of torques show a dramatic enhancement at large bias voltage (iii) the out-of-plane torque dominates in-plane torque at large biases, (iv) at very high bias voltages, the TMR and spin torque shows an oscillatory dependence. Finally, we discussed the experimental conditions of the observation of current-driven magnetic domain wall motion and magnetization switching.

\section{Acknowledgement}
The authors gratefully acknowledge insightful discussions with J.S. Moodera, G.X. Miao and B. Hillebrands.


\newpage
\addcontentsline{toc}{section}{8 Bibliography}
\begin{thebibliography}{999}
\bibitem{seebeck} K. Uchida, J. Xiao, H. Adachi, J. Ohe, S. Takahashi, J. Ieda, T. Ota, Y. Kajiwara, H. Umezawa, H. Kawai, G. E.W. Bauer, S. Maekawa and E. Saitoh, Nature Materials {\bf9}, 894 (2010).
\bibitem{hillebrand} See e.g., S.O. Demokritov, A. A. Serga, A. Andre,V. E. Demidov, M. P. Kostylev, B. Hillebrands, and A. N. Slavin, Phys. Rev. Lett. {\bf93}, 047201 (2004); A. A. Serga, S.O. Demokritov, B. Hillebrands, and A. N. Slavin, Phys. Rev. Lett. {\bf92}, 117203 (2004); U.-H. Hansen, M. Gatzen, V. E. Demidov, and S. O. Demokritov, Phys. Rev. Lett. {\bf99}, 127204 (2007); S. O. Demokritov, V. E. Demidov, O. Dzyapko, G. A. Melkov, A. A. Serga, B. Hillebrands, and A. N. Slavin, Nature {\bf443}, 430 (2006).
\bibitem{Kajiwara} Y. Kajiwara, K. Harii, S. Takahashi, J. Ohe, K. Uchida, M. Mizuguchi, H. Umezawa, H. Kawai, K. Ando, K. Takanashi, S. Maekawa and E. Saitoh, Nature {\bf 464}, 262 (2010).
\bibitem{wang} Z. Wang, Y. Sun, M. Wu, V. Tiberkevich, and A. Slavin, Phys. Rev. Lett. {\bf107}, 146602 (2011); Z. Wang, Y. Sun, Y.-Y. Song, M. Wu, H. Schulthei, J. E. Pearson, and A. Hoffmann, Appl. Phys. Lett. {\bf99}, 162511 (2011).
\bibitem{Slon1996} J.C. Slonczewski, J. Magn. Magn. Mater. {\bf 159}, L1 (1996); L.Berger, Phys. Rev. B {\bf 54},9353 (1996).
\bibitem{tserko} Y. Tserkovnyak, A. Brataas and G. E. W. Bauer, Phys. Rev. Lett. {\bf88}, 117601 (2002); Phys. Rev. B {\bf66}, 224403 (2002).
\bibitem{sandweg} C. W. Sandweg, Y. Kajiwara, K. Ando, E. Saitoh, and B. Hillebrands, Appl. Phys. Lett. {\bf97}, 252504 (2010);
C. Burrowes, B. Heinrich, B. Kardasz, E. A. Montoya, E. Girt, Y. Sun, Y.-Y. Song, and M. Wu, Appl. Phys. Lett. {\bf100}, 092403 (2012).
\bibitem{heinrich} B. Heinrich, C. Burrowes, E. Montoya, B. Kardasz, E. Girt, Y.Y. Song, Y. Sun, and M. Wu, Phys. Rev. Lett. {\bf107}, 066604 (2011).
\bibitem{stiles2002} M.D. Stiles and A. Zangwill, Phys. Rev. B {\bf66}, 014407 (2002).
\bibitem{slonc2010} J. C. Slonczewski Phys. Rev. B {\bf82}, 054403 (2010).
\bibitem{xiao} J. Xiao, G. E. W. Bauer, K.C. Uchida, E. Saitoh, and S. Maekawa, Phys. Rev. B {\bf81}, 214418 (2010).
\bibitem{moodera95} J. S. Moodera, L.R. Kinder, T.M. Wong, and R. Meservey,, Phys. Rev. Lett. {\bf 74}, 3273 (1995).
\bibitem{mgoth} W. H. Butler, X.G. Zhang, T.C. Schulthess, and J. M. MacLaren, Phys.Rev. B {\bf 63}, 054416 (2001).
\bibitem{yuasa} S. Yuasa, A. Fukushima, T. Nagahama, K. Ando, and Y. Suzuki, Nat. Mater. {\bf3}, 868 (2004); S. S. P. Parkin, C. Kaiser, A. Panchula, P. M. Rice, B. Hughes, M. Samant, and S. H. Yang, Nat. Mater. {\bf3}, 862 (2004).
\bibitem{ikeda:082508} S. Ikeda J. Hayakawa, Y. Ashizawa, Y. M. Lee, K. Miura, H. Hasegawa, M. Tsunoda, F. Matsukura, and H. Ohno, Appl. Phys. Lett. {\bf 93}, 082508 (2008).
\bibitem{jiangmgo} L. X. Jiang, H. Naganuma, M. Oogane, and Y. Ando, Appl. Phys. Express {\bf 2}, 083002 (2009).
\bibitem{spinfilter} J. S. Moodera, T. S. Santos and T. Nagahama, J. Phys.: Condens. Matter {\bf19}, 165202 (2007).
\bibitem{nagahama} T. Nagahama, T.S. Santos, and J.S. Moodera, Phys. Rev. Lett. {\bf 99}, 016602 (2007).
\bibitem{leclair} P. LeClair \textit{et al}., Appl. Phys. Lett. {\bf 80}, 625 (2002).
\bibitem{moodera88} J. S. Moodera, X. Hao, G.A. Gibson and R. Meservey Phys. Rev. Lett. 61, 637 (1988); X. Hao, J. S. Moodera and R. Meservey, Phys. Rev. B {\bf42} 8235 (1990).
\bibitem{NiFe2O4} U. Luders \textit{et al}., Appl. Phys. Lett. {\bf 88}, 082505 (2006); % single spin filtering junction NiFe2O4
\bibitem{CoFe2O4} A.V. Ramos \textit{et al}., Appl. Phys. Lett. {\bf 91}, 122107 (2007).%single spin filter CoFe2O4
\bibitem{BiMnO3} M. Gajek \textit{et al}., Phys. Rev. B {\bf 72}, 020406(R) (2005).%single spin filter BiMnO3
\bibitem{SantosEuO} T. S. Santos \textit{et al}., Phys. Rev. Lett. {\bf 101}, 147201 (2008).
\bibitem{Miao} G.X. Miao, M. Muller, and J.S. Moodera, Phys. Rev. Lett. {\bf 102}, 076601 (2009).
\bibitem{miaojap} G.X. Miao and J.S. Moodera, J. Appl. Phys. {\bf 106}, 023911 (2009).
\bibitem{reviewtsymbal} E. Y. Tsymbal, O.N. Mryasov and P. R. LeClair, J. Phys.: Condens. Matter {\bf15}, R109-R142 (2003).
\bibitem{inoue} J.I. Inoue, Phys. Rev. B {\bf84}, 180402(R) (2011).
\bibitem{jia} X. Jia, K. Liu, K. Xia and G. E. W. Bauer, EuroPhys. Lett., {\bf 96} 17005 (2011).
\bibitem{Shishidou} T. Shishidou, N. Mikamo, Y. Uratani, F. Ishii, and T. Oguchi, J. Phys. Cond. Mat. 16, S5677 (2004).
\bibitem{Szotek} Z. Szotek, W. M. Temmerman, D. Kodderitzsch, A. Svane, L. Petit, and H. Winter, Phys. Rev. B {\bf 74}, (174431) (2006).
\bibitem{Stiles} M. D. Stiles and A. Zangwill, Phys. Rev. B {\bf 66}, 014407 (2002).
\bibitem{Ralph} D. C. Ralph and M. D. Stiles, J. Magn. Magn. Mater. {\bf 320}, 7, 1190 (2008).
\bibitem{Fowler} R. H. Fowler and L. Nordheim, Proc. R. Soc. London, {\bf 119}, 173(1928).
\bibitem{montaigne} F. Montaigne, M. Hehn, and A. Schuhl, Phys. Rev. B {\bf64}, 144402 (2001).
\bibitem{Theodonis} I. Theodonis, N. Kioussis, A. Kalitsov, M. Chshiev, and W. H. Butler, Phys.Rev. Lett. {\bf97}, 237205 (2006); A. Kalitsov, M. Chshiev, I. Theodonis, N. Kioussis, and W. H. Butler, Phys. Rev. B {\bf79}, 174416 (2009).
\bibitem{manchon2008} A. Manchon, N. Ryzhanova, A. Vedyayev, M. Chschiev and
B. Dieny, J. Phys.: Cond. Matt. {\bf20}, 145208 (2008).
\bibitem{Heiliger} C. Heiliger and M.D. Stiles, Phys. Rev. Lett. {\bf 100}, 186805 (2008).
\bibitem{Tang} Y.-H. Tang, N. Kioussis, A. Kalitsov, W. H. Butler, and R. Car, Phys. Rev. Lett. {\bf103}, 057206 (2009); Phys. Rev. B {\bf81}, 054437 (2010).
\bibitem{wil} M. Wilczynski, J. Barnas, and R. Swirkowicz, Phys. Rev. B {\bf77}, 054434 (2008).
\bibitem{Huai} J.Z. Sun, J. Magn. Magn. Mater. {\bf202}, 157 (1999); Y. Huai, F. Albert, P. Nguyen, M. Pakala, and T. Valet, Appl. Phys. Lett. {\bf84}, 3118
(2004); G. D. Fuchs, N. C. Emley, I. N. Krivorotov, P. M.
Braganca, E. M. Ryan, S. I. Kiselev, J. C. Sankey, D. C. Ralph, R.
A. Buhrman, and J. A. Katine, Appl. Phys. Lett. {\bf85}, 1205
(2004); D. Chiba, Y. Sato, T. Kita, F. Matsukura, and H. Ohno, Phys. Rev. Lett. {\bf93}, 216602 (2004).
%\bibitem{oscillator}D. Houssameddine \textit{et al}., Nat. Mater. {\bf 6}, 447 (2007).
\bibitem{Ikeda} S. Ikeda \textit{et al}., IEEE Transactions on electron devices {\bf 54}, 991 (2007).
\bibitem{Chanthbouala} A. Chanthbouala, R. Matsumoto, J. Grollier, V. Cros, A. Anane, A. Fert, A. V. Khvalkovskiy,
K. A. Zvezdin, K. Nishimura, Y. Nagamine, H. Maehara, K. Tsunekawa, A. Fukushima and S. Yuasa, Nature Physics {\bf7}, 626 (2011).
\bibitem{Sankey} J. C. Sankey, Y.-T. Cui, R. A. Buhrman, D. C. Ralph, J. Z. Sun, and J. C. Slonczewski, Nature Physics {\bf4}, 67
(2008); H. Kubota, A. Fukushima, K. Yakushiji, T. Nagahama, S. Yuasa, K. Ando, H. Maehara, Y. Nagamine, K. Tsunekawa, D. D. Djayaprawira, N. Watanabe, and Y. Suzuki, Nature Physics {\bf4}, 37 (2008).
\bibitem{notling} R. Schiller and W. Nolting, Phys. Rev. Lett. {\bf86}, 3847 (2001).
\end{thebibliography}
\end{document}